\documentstyle[11pt,fleqn,epsf,a4]{article}

\newcommand{\Tr}{\makebox{ Tr }}

\newcommand{\fm}{\makebox{ fm}}

\newcommand{\beq}{\begin{equation}}

\newcommand{\enq}{\end{equation}}

\newcommand{\beqa}{\begin{eqnarray}}

\newcommand{\enqa}{\end{eqnarray}}

\newcommand{\nn}{\nonumber}

\newcommand{\lbq}[1]{\label{#1} \enq}

\newcommand{\lbqa}[1]{\label{#1} \enqa}

\newcommand{\befi}[1]{\begin{figure}[ht] \leavevmode \centering
\epsffile{#1.eps}}

\newcommand{\lbfi}[1]{\label{#1} \end{figure}\parindent0em}

\newcommand{\eq}[1]{eq.(\ref{#1})}

\newcommand{\fig}[1]{fig.(\ref{#1})}

\newcommand{\ct}{\cite}

\newcommand{\pa}{\partial}

\newcommand{\cD}{\mbox{$\cal D$}}

\newcommand{\cP}{\mbox{$\cal P$}}

\newcommand{\cN}{\mbox{$\cal N$}}

\newcommand{\bA}{\mbox{\bf A}}

\newcommand{\bD}{\mbox{\bf D}}

\newcommand{\bF}{\mbox{\bf F}}

\newcommand{\al}{\alpha}

\newcommand{\be}{\beta}

\newcommand{\ga}{\gamma}

\newcommand{\de}{\delta}

\newcommand{\ka}{\kappa}

\newcommand{\la}{\lambda}

\newcommand{\rh}{\rho}

\newcommand{\Th}{\Theta}

\begin{document}
\sloppy \title{\begin{flushright}\normalsize
  HD-THEP-95-12\end{flushright} \LARGE \bf STRING FORMATION IN THE
  MODEL OF THE STOCHASTIC VACUUM AND CONSISTENCY WITH LOW ENERGY
  THEOREMS} \author{ H. G. Dosch, O. Nachtmann, Michael
  Rueter\thanks{supported by the Deutsche
    Forschungsgemeinschaft}\\[.7cm] \it Institut f\"ur theoretische
  Physik\\ \it Universit\"at Heidelberg\\ \it Philosophenweg 16,
  D-69120 Heidelberg, FRG\\[.3cm] \it e-mail:
  rueter@next3.thphys.uni-heidelberg.de } \date{} \maketitle
\thispagestyle{empty}
\begin{abstract}
  We re-derive and discuss two low energy theorems which relate the
  potential of a static quark-antiquark pair with the total energy and
  action stored in the flux tube between the sources. In lattice QCD
  these relations are known as Michael's sum-rules, but we give an
  essential correction to one of them. Then we relate the low energy
  theorems to the virial theorem for a heavy quark-antiquark bound
  state. Finally we compare the results for the flux tube formation,
  which have been calculated in the model of the stochastic vacuum,
  with the low energy theorems and obtain consistency. From this we
  conclude that the model describes the non perturbative gluon
  dynamics of QCD at a renormalization scale, where the strong
  coupling constant is given by $\al_s= 0.57$.
\end{abstract}
\newpage \setcounter{page}{1}
\section{Introduction}

As it is well known there is no analytical tool comparable to
perturbation theory for treating QCD processes involving large
distances. The most direct approach is a numerical evaluation of the
lattice regularized version of QCD. Nevertheless it seems useful also
to try analytical approaches, based on model assumptions, which may
reveal some of the underlying structure of the theory. The success of
the QCD sum-rules makes it plausible that the properties of the
nontrivial QCD-vacuum play an essential role for the infra-red
behavior of the theory.\\ The model of the stochastic vacuum (MSV)
makes the crucial assumption that the complicated infra-red behavior
of non-Abelian gauge theories can be adequately approximated by a
cluster expansion of the stochastic process describing the quantized
theory \ct{1}\ct{2}\ct{3}. It has been shown that such a model leads
to linear confinement. In order to make more specific and quantitative
predictions one has to make more radical assumptions, the most radical
one being that all higher cumulants can be neglected as compared to
the two point function. Then the stochastic process is a Gaussian one,
i.e.~all higher correlators can be reduced to products of two point
functions by factorization. We refer to \ct{3}\ct{4} for the
description of the model and its applications. In this note we present
some consistency checks by comparing results of the model for heavy
quark-antiquark systems with low energy theorems.

Our paper is organized as follows: In section \ref{sec:1} we re-derive
some low energy theorems, in section \ref{sec:2} we show how these
theorems relate to the virial theorem for a heavy quark-antiquark
bound state. In section \ref{sec:3} we present the results for the
string formation calculated in the MSV and in section \ref{sec:4} we
compare these findings with the low energy theorems and discuss them.

Except for section \ref{sec:2} we work with an Euclidean field theory
(EFT), where the square of the electric field has the opposite sign of
the same quantity in a Minkowski field theory (MFT), whereas the
square of the magnetic field has the same sign in both theories:
\[  \vec E^2_{\rm EFT} = - \vec E^2_{\rm MFT}, \quad
\vec B^2_{\rm EFT} = \vec B^2_{\rm MFT}\; .
\]
Here and in the following a sum over the $N_c^2-1$ color components is
understood.
\section{The Low Energy Theorems}
\label{sec:1}

In this section we consider pure gluon QCD and static quark-antiquark
sources. We follow the procedure of Novikov, Shifman, Vainshtein and
Zakharov (NSVZ) \ct{N1} and redefine the unrenormalized matrix valued
gluon field tensor by multiplying it with the bare coupling constant
$g_0$: \beq \bar {\bF}_{\mu \nu} \equiv g_0 \bF^{(0)}_{\mu \nu} \; .
\lbq{1} The QCD-action $S_{QCD}$ then depends on
$\al_{s0}=\frac{g_0^2}{4\pi}$ only in the form \beq S_{QCD}\equiv
\frac{1}{8 \pi \al_{s0}}\int {\rm d^4}x\Tr [\bar {\bF}_{\mu
  \nu}(x)\bar{ \bF}_{\mu \nu}(x)] \; .  \enq Let us consider the
vacuum expectation value of a Wegner-Wilson loop \ct{N8}\ct{N9}: \beq
<W[L]> \equiv <\Tr \cP \exp[-i\int_L \bar{ \bA}_\mu d x_\mu]> \;
.\lbq{2} Here $L$ denotes a closed loop and $\cal{P}$ path ordering.
In a somewhat symbolic notation we write this as a functional
integral: \beqa <W[L]> &=& \frac{1} {\cN} \int \cD A \exp[-S_{QCD}]
\Tr \cP \exp[-i\int_L \bar{ \bA}_\mu d x_\mu]\; ,\nn \\ \cN&\equiv
&\int \cD A \exp[-S_{QCD}]\; .  \lbqa{3} Differentiating $\log <W[L]>$
with respect to $-\frac{1}{8\pi \al_{s0}}$ we obtain: \beqa && 8\pi
\al_{s0}^2 \frac{\pa\log[\, <W[L]>\, ]}{\pa \al_{s0}} = \nn \\ &&
\frac{1}{<W[L]>} <\int {\rm d^4}x \Tr [\bar{ \bF}_{\mu \nu}(x) \bar{
  \bF}_{\mu \nu}(x)]\Tr \cP \exp[-i\int_L \bar{ \bA}_\mu d x_\mu]> \nn
\\ &&- <\int {\rm d^4}x \Tr [\bar{ \bF}_{\mu \nu}(x)\bar{ \bF}_{\mu
  \nu}(x)]>\; .  \lbqa{4} If the loop $L$ is a rectangle with
``spatial'' extension $-R/2<x_3<R/2$ and ``temporal'' extension $-T/2
< x_4 < T/2 $ the right-hand side (rhs) of \eq{4} can be simplified
further in the limit of large $T$. Let the length $a$ be of the order
of the correlation length of the gluon field strengths. Then for
$|x_4|> T/2 + a$ the rhs of \eq{4} is zero and for $|x_4| < T/2 -a$
independent of $x_4$. This gives for \eq{4} in the limit of large $T$:
\beq 8\pi \al_{s0}^2 \frac{\pa\log[\, <W[L]>\, ]}{\pa \al_{s0}} = T <
\int {\rm d^3}x \Tr [\bar{ \bF}_{\mu \nu}(x)\bar {\bF}_{\mu
  \nu}(x)]>_R \; .\lbq{5} Here and in the following $x_4$ is zero and
the expectation value $< \; . \; >_R$ shall denote the EFT expectation
value in the presence of a static quark-antiquark pair at distance
$R$, i.e.~static sources in the fundamental representation, where the
expectation value in absence of the sources has been subtracted.

We define the potential $V(R)$ by: \beq V(R) \equiv \lim_{T\to
  \infty}- \frac{\log <W[L]>}{T} \lbq{6} and thus obtain the low
energy theorem: \beq -4\pi\al_{s0}^2 \frac{\pa V(R)}{\pa \al_{s0}} =
\frac{1}{2}< \int {\rm d^3}x\Tr [\bar{ \bF}_{\mu \nu}(x)\bar{
  \bF}_{\mu \nu}(x)]>_R \; .\lbq{7} In \eq{6} and \eq{7} self energies
are understood to be subtracted.

Up to now we have used the bare coupling $g_0$ and bare fields. If we
use the background gauge fixing \ct{N2} we have \beq g=Z_g^{-1}g_0\; ,
\; \bA_\mu = Z^{-1/2}\bA _\mu^{(0)}\; , \; Z_g Z^{1/2}=1\; , \; \bar{
  \bA}_\mu = g_0 \bA^{(0)}_\mu = g \bA_\mu\; , \lbq{gauge} where $g$
and $\bA_\mu$ are renormalized quantities. The renormalized squared
field strength tensor is given by: \beq \Tr [\bF_{\mu \nu}(x)\bF_{\mu
  \nu}(x)]=Z_{F^2}^{-1}\Tr [\bF^{(0)}_{\mu \nu}(x)\bF^{(0)}_{\mu
  \nu}(x)]\; , \lbq{gauge4} where $Z_{F^2}$ was calculated \ct{N6} in
background gauge with Landau gauge fixing to be
\[ Z_{F^2}=1+\frac{1}{2}g_0\frac{\pa}{\pa g_0} \log Z \]
and thus we obtain \beq \frac{\pa \al_s}{\pa \al_{s0}}= Z Z_{F^2}\; .
\lbq{gauge3} We use \eq{gauge}-\eq{gauge3} to express \eq{7} in terms
of renormalized quantities: \beq -\al_{s}\frac{\pa V(R)}{\pa \al_{s}}
= \frac{1}{2}< \int {\rm d^3}x\Tr [\bF_{\mu \nu}(x)\bF_{\mu\nu}(x)]>_R
\; .\lbq{7a} In order to evaluate the left-hand side (lhs) of \eq{7a}
we use standard renormalization group arguments: The only scales are
the renormalization scale $\mu$ and $R$, thus the potential $V=V(R,\mu
,\al_S)$ satisfies on dimensional grounds: \beq \left( R\frac{\pa}{\pa
  R}- \mu\frac{\pa}{\pa \mu}\right)V(R,\mu ,\al_s)=- V(R,\mu ,\al_s)\;
.  \lbq{gauge5} But for $V$ as a physical quantity the total
derivative with respect to $\mu$ must vanish: \beq \mu\frac{\rm d}{\rm
  d\mu} V(R, \mu, \al_{s}) = \left( \mu\frac{\pa}{\pa \mu}+
\tilde{\be}(\al_s)\frac{\pa}{\pa \al_s}\right)V(R,\mu ,\al_s)= 0\; ,
\lbq{8} where $\tilde{\be}(\al_{s}) \equiv \mu \frac{\rm d}{\rm d\mu}
\al_{s}(\mu)$. We get from \eq{gauge5} and \eq{8}: \beq \frac{\pa
  V(R)}{\pa\al_{s}} = -\frac{1}{\tilde{\be}} \left\{ V(R) + R
\frac{\pa V(R)}{\pa R} \right\} \lbq{9} and we finally obtain with
\eq{7a} the low energy theorem \beq \left\{ V(R) + R \frac{\pa
  V(R)}{\pa R} \right\} = \frac{1}{2} \frac{\tilde{\be}}{\al_{s}} <
\int {\rm d^3}x\Tr [\bF_{\mu \nu}(x)\bF_{\mu\nu}(x)]>_R \lbq{10} or
equivalently \beq \left\{ V(R) + R \frac{\pa V(R)}{\pa R}
\right\}=\frac{1}{2}\frac{\tilde{\be}}{\al_{s}}< \int {\rm
  d^3}x\left(\vec{E}(x)^2+\vec{B}(x)^2 \right)>_R\; .  \lbq{11} The
\eq{11} is for the case of a linear potential just one of the many low
energy theorems derived by NSVZ \ct{N1}. But there renormalization was
only discussed to leading order.

We have also the usual relation between the potential $V(R)$ and the
energy density $\Th_{00}(x)$: \beq V(R) = \int {\rm d^3} x
<\Th_{00}(x)>_R = \frac{1}{2}< \int {\rm d^3} x \left( -\vec E(x)^2 +
\vec B(x)^2 \right)>_R\; .  \lbq{12} Note that we are in a Euclidean
field theory, hence the minus sign of the electric field, and that on
the rhs of \eq{11} and \eq{12} we have renormalized composite
operators.

In lattice QCD \eq{11} and \eq{12} are known as Michael's sum-rules
\ct{N7}. But in the derivation of the action sum-rule (\eq{11})
scaling of $V(R)$ with $R$ was not taken properly into account and
hence the second term on the lhs of \eq{11} is missing in \ct{N7}.
\section{The Low Energy Theorems For A Quark-Antiquark Bound State}
\label{sec:2}

In this section we work in Minkowski field theory of gluons
interacting with a dynamical heavy quark field $q$. The symmetrized
energy momentum tensor is given by: \beqa \Th_{\ka\la}^{\rm sym.} &=&
\Th_{\ka\la}^{\rm G}+\Th_{\ka\la}^{\rm Q}\; ,\nn \\ \Th_{\ka\la}^{\rm
  G} &= & 2\Tr [\bF_{\ka\rh} {\bF^{\rh}}_\la ]+\frac{1}{2}
g_{\ka\la}\Tr [\bF_{\mu\nu}\bF^{\mu\nu}]\; , \nn \\ \Th_{\ka\la}^{\rm
  Q} &=& \frac{i}{4} \bar{q} \left( \ga_\ka
\stackrel{\leftrightarrow}{\bD}_\la +\ga_\la
\stackrel{\leftrightarrow}{\bD}_\ka \right) q\; .  \lbqa{tr1} Here
$\bD_\la$ is the covariant derivative.

Consider a quark-antiquark bound state, which is an eigenstate of
energy and momentum: \beqa |M(p)> &=& |q\bar{q}(p)> \; ,\nn \\
<M(p')|M(p)> &=& 2p^0 (2\pi)^3 \de^3 (\vec{p}\, '-\vec{p}\, )\; .
\lbqa{tr2} We have from Lorentz invariance \beq
<M(p)|\Th_{\mu\nu}^{\rm sym.}(0)|M(p)>= 2 p_\mu p_\nu \; .  \lbq{tr3}
In the rest frame of the bound state the total energy is given by
$p^0_R=2m_q+E_B$, where the binding energy $E_B$ is the sum of the
mean values of the potential and kinetic energies: \beq E_B= <V>+<T>\;
.  \lbq{tr4} With \eq{tr3} and $\Th^{\rm sym.}\equiv
g^{\mu\nu}\Th_{\mu\nu}^{\rm sym.}$ we obtain for $|E_B| \ll m_q$:
\beqa <M(p_R)|\Th_{00}^{\rm sym.}(0)|M(p_R)> &=& <M(p_R)|\Th^{\rm
  sym.}(0)|M(p_R)> \nn \\ &=& 8m_q^2+8m_qE_B\; .  \lbqa{tr5} The
kinetic part of the energy momentum tensor of the quark-antiquark
bound state is $\Th_{\mu\nu}^{\rm Q}$. With the kinetic momenta of the
quarks, $(p_q)^2=(p_{\bar{q}} )^2=m_q^2$, we have \beq
\frac{1}{p_R^0}<M(p_R)|\Th_{\mu\nu}^{\rm Q}(0)|M(p_R)>=<\frac{2p_{q\,
    \mu}p_{q\, \nu}}{p_q^0}+\frac{2p_{\bar{q}\, \mu}p_{\bar{q}\,
    \nu}}{p_{\bar{q}}^0}> \; .  \lbq{tr6} This gives for $\Th^{\rm
  Q}\equiv g^{\mu\nu}\Th_{\mu\nu}^{\rm Q}$ and $\Th_{00}^{\rm Q}$:
\beqa \frac{1}{2p_R^0}<M(p_R)|\Th^{\rm Q}(0)|M(p_R)> &=& 2m_q-<T>\; ,
\nn \\ \frac{1}{2p_R^0}<M(p_R)|\Th_{00}^{\rm Q}(0)|M(p_R)>&=&
2m_q+<T>\; .  \lbqa{tr7} More details on the derivation of \eq{tr6}
and \eq{tr7} will be given in a future publication. Now we evaluate
the gluonic part of the energy momentum tensor by subtracting the
kinetic part (\eq{tr7}) from the total tensor (\eq{tr5}). We obtain
\beqa \frac{1}{2p_R^0}<M(p_R)|\Th^{\rm G}(0)|M(p_R)> &=& <V>+2<T> \; ,
\nn \\ \frac{1}{2p_R^0}<M(p_R)|\Th_{00}^{\rm G}(0)|M(p_R)>&=&<V>\;
\lbqa{tr8} with $\Th^{\rm G}\equiv g^{\mu\nu}\Th_{\mu\nu}^{\rm G}$. In
order to relate the kinetic energy to the potential one we use the
virial theorem:
\[ <T>=\frac{1}{2}<R\; \frac{\pa V}{\pa R}>\; . \]
Eq.(\ref{tr8}) then gives \beqa \frac{1}{2p_R^0}<M(p_R)|\Th^{\rm
  G}(0)|M(p_R)> &=& <V>+<R\; \frac{\pa V}{\pa R}> \; , \nn \\
\frac{1}{2p_R^0}<M(p_R)|\Th_{00}^{\rm G}(0)|M(p_R)>&=& <V>\; .
\lbqa{tr9} Using the trace anomaly of the energy momentum tensor
\ct{N10}
\[ \Th^{\rm G} = \frac{\tilde{\be}}{2\al_s}\Tr [ \bF_{\mu\nu}
\bF^{\mu\nu} ]\]
we finally obtain: {\mathindent-1em \beqa <V>+<R\; \frac{\pa V}{\pa
    R}> &=& \frac{1}{2p_R^0}<M(p_R)|\frac{\tilde{\be}}{2\al_s}\Tr [
  \bF_{\mu\nu}(0)\bF^{\mu\nu}(0)]|M(p_R)> \; , \label{tr10a}\\ <V> &=&
  \frac{1}{2p_R^0}<M(p_R)|\Th_{00}^{\rm G}(0)|M(p_R)>\; .
  \lbqa{tr10}}\\[0em] It is straightforward to show that here we have
again the relations (\ref{10}) and (\ref{12}), but in \eq{tr10a} and
\eq{tr10} we have mean values as obtained by integration with the
absolute square of the wave function of the quark-antiquark bound
state. The integration over the space points in \eq{10} and \eq{12}
corresponds to the translational invariance of the quark-antiquark
bound state in \eq{tr10a} and \eq{tr10}.
\section{The Flux Tube In The MSV}
\label{sec:3}

In this section we present the main results from a calculation of the
flux tube between a static quark-antiquark pair in the MSV. For all
technical details we refer to our previous publication \ct{N3}.

The expectation values of the squared elements of the gluon field
strength tensor in presence of the quark-antiquark pair minus the
vacuum values are given by: \beq \Delta F^2_{\alpha\beta}\:(x)\;
\equiv \; \frac{4}{R_P^4} \frac{<W[L]\:
  P_{\alpha\beta}(x)>-<W[L]><P_{\alpha\beta}(x)>}{<W[L]>}\; .
\lbq{21} Here $P_{\alpha\beta}(x)$ is a Wegner-Wilson loop over a
plaquette in $\al,\be$-direction centered at $x$ with side length
$R_P$. For $R_P\rightarrow 0$ we have: \beq \lim_{R_P\to 0}\Delta
F^2_{\alpha\beta}\:(x) = -2<g^2\Tr [\bF_{\al\be}(x)\bF_{\al\be}(x)]>_R
\enq where there is no summation over $\al$ and $\be$.

\begin{figure}[ht]
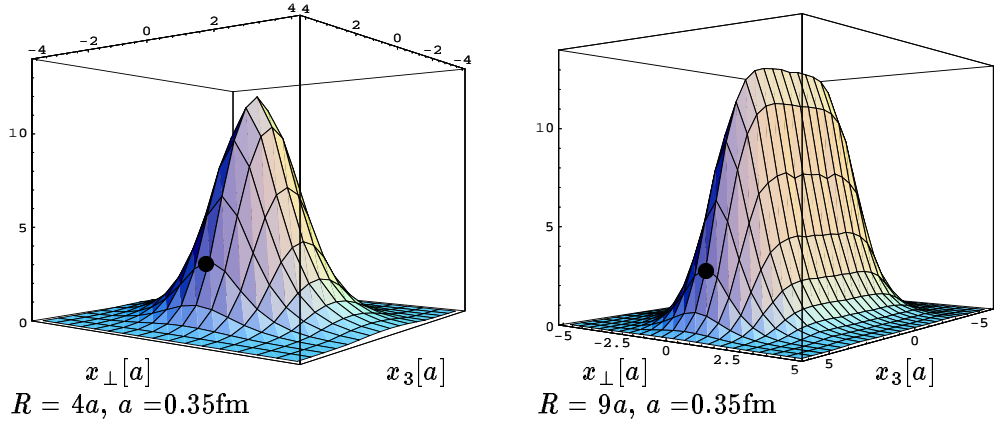

\begin{minipage}{6.2cm}
  \epsfysize5cm \epsfxsize6.2cm
  $R=4a$, $a=$0.35fm
\end{minipage}
\hfill
\begin{minipage}{6.2cm}
  \epsfysize5cm \epsfxsize6.2cm
  $R=9a$, $a=$0.35fm
\end{minipage}
\caption{Difference of the squared electric field parallel to the
quark-antiquark ($x_3$) axis \mbox{($-<g^2E_\|^2(x_3,x_\perp )>_R$)}
in $\rm \frac{GeV}{fm^3}$ for different quark separations $R$.
$x_\perp$ is the distance transverse to the $x_3$-axis and the dots
denote the quark positions.}
\label{34c}
\end{figure}

It was shown in \ct{N3} that by symmetry arguments
$<g^2\vec{B}^2>_R=0$. The square of the electric field perpendicular
to the loop $L$ ($<g^2E_\perp ^2>$) is also practically not influenced
but only the squared electric field parallel to $L$ ($<g^2 E_\| ^2>$)
is affected by the static sources. In \fig{34c} we display $-<g^2 E_\|
^2>_R$ as a function of the perpendicular distance $x_\perp$ from the
loop $L$ and of the position along the quark-antiquark axis $x_3$. The
lengths are given in units of the correlation length $a$ of the gluon
field strengths, which is fixed to $a=0.35\fm$ in the MSV \ct{N3}.

\parindent0em

\begin{figure}[ht]
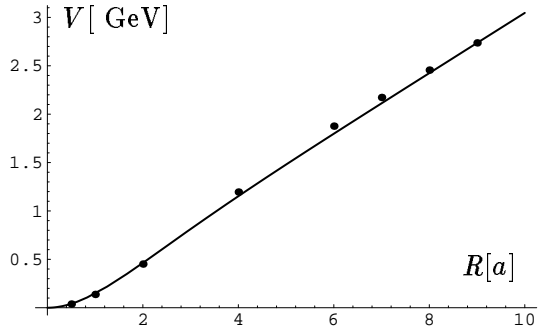

\caption{The total energy stored in the field (dots) calculated with
the energy sum-rule (\protect\ref{12}) as compared with the potential
of a q-$\rm \bar{q}$-pair obtained by evaluation of the Wegner-Wilson
loop in the MSV \protect\ct{1}\protect\ct{2} (solid line)}
\label{Egesamt}
\end{figure}

As can be seen directly from \eq{11} and \eq{12} the squared electric
and magnetic field strengths are separately not renormalization group
invariant, thus statements like $<g^2\vec{B}^2>_R=0$ or $<g^2E^2_\perp
>_R\approx 0$ are scale dependent. In \ct{N3} we have used \eq{12} in
order to determine at which $\al_s$ the MSV is supposed to work. We
have calculated the total energy stored in the flux tube and compared
it to the potential $V(R)$, which can be determined directly in the
MSV \ct{1}\ct{2}. In \fig{Egesamt} we present this comparison using
our fitted value $\al_s=0.57$ \ct{N3}.  \epsfxsize7cm
\parindent0em
\section{Discussion}
\label{sec:4}

We first consider the region where the potential is linear. Then
\eq{11} reads: \beq \frac{1}{2}\frac{\tilde{\be}}{\al_{s}}< \int {\rm
  d^3}x\left(\vec{E}(x)^2+\vec{B}(x)^2 \right)>_R = 2\;V(R) \;
.\lbq{31} Since in the model of the stochastic vacuum the static
sources do not modify the color magnetic field, i.e.~$<\vec B^2>_R =
0$ , we have consistency between \eq{12} and \eq{31} only if \beq
\tilde{\be}/\al_s = -2\; . \lbq{32} From the energy sum-rule (\eq{12})
we have already obtained $\al_s=0.57$, thus consistency of the MSV
requires that \eq{32} should be satisfied for this value of $\al_s$.

The $\tilde{\be}$-function can be expanded in a power series in
$\al_s$: \beq \tilde{\be}(\al_s) = -\frac{\be_0}{2 \pi}\al_s^2
-\frac{\be_1}{4 \pi^2}\al_s^3 -\frac{\be_2}{64 \pi^3} \al_s^4+ \ldots
\; .\lbq{33} The n-loop expansion gives the series up to $\al_s^{\rm
  n+1}$. The first two terms of the power series are scheme
independent, the numerical value of the term proportional to $\al_s^4$
has been evaluated \ct{N11} in $\overline{MS}$. For pure gauge SU(3)
theory we have \ct{N4}: \beq \be_0=11, \quad \be_1 = 51, \quad
{\be_2}_{\overline{MS}} = 2857 \; .\lbq{34} Numerical investigations
of the running coupling constant in lattice gauge Monte Carlo
simulations \ct{N5} indicate that the power series (\ref{33}) for
$\tilde{\be}(\al_s)$ truncated at $\al_s^4$ can be used up to rather
large values of $\al_s$ in their scheme if the coefficient $\be_2$ is
fitted numerically with the result $\be_2\approx 6270$.

{}From \fig{btildeOVERalpha} we see that our condition (\ref{32}) yields
$\al_s=1.14$ using $\tilde{\be}$ on the one loop level, $\al_s=0.74$
on the two loop level and $\al_s=0.64$ on the three loop level in
$\overline{MS}$. This value is already very close to the value of
$\al_s=0.57$ determined from the flux tube calculation using \eq{12}.
If we use for $\tilde{\be}$ the series (\ref{33}) truncated at order
$\al_s^4$ and impose condition (\ref{32}) for $\al_s=0.57$ and
determine the coefficient $\be_2$ accordingly we find $\be_2=6250$
(see \fig{btildeOVERalpha}), i.e.~about twice the $\overline{MS}$
value. Curiously this agrees almost exactly with the numerically
determined value of \ct{N5} quoted above.

\epsfxsize7cm
\begin{figure}[ht]
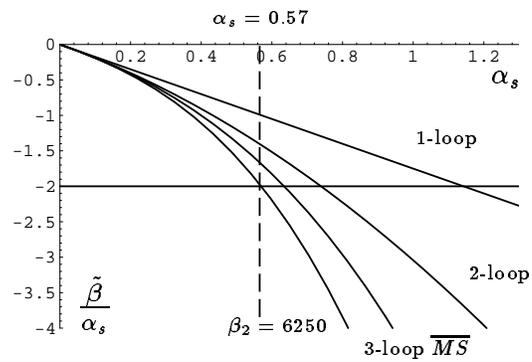
\caption{$\frac{\tilde{\be}}{\al_s}$ as a function of $\al_s$ for
different $\tilde{\be}$-functions}
\lbfi{btildeOVERalpha}

We see from \fig{Egesamt} that for distances $R/a \leq 1\; (a=0.35
\fm)$ the MSV does no longer yield a linearly rising potential and
hence we have
\[ V(R) + R\frac{\pa V(R)}{\pa R} \neq 2 V(R)\; , \qquad R/a
\leq 1\; . \]
Therefore the conditions (\ref{12}) and (\ref{31}) cannot be fulfilled
simultaneously for $<\vec B^2>_R =0$. This does by no means speak
against the model, since in the derivation of the results in \ct{N3}
it was stressed that the spatial extension of the loop had to be large
as compared to the correlation length of the field strengths in order
to justify the evaluation of $\vec E^2$ and $\vec B^2$.
\section{Conclusions}

We have re-derived and discussed low energy theorems and sum-rules
relating densities of squared field strengths to the potential of a
static quark-antiquark pair. We found that for a dynamic
quark-antiquark bound state these sum-rules are in a sense equivalent
to the dilatation trace anomaly and the virial theorem. We have
applied the sum-rules to quantities obtained in the model of the
stochastic vacuum. This model was introduced as a phenomenological
approximation to the complicated measure for functional integration
over the slowly varying gluon field strengths. No use of the equations
of motion has been made explicitly. The factorization hypothesis
(Gaussian measure) enters crucially in the evaluation of the squared
color fields. In view of the drastic approximations made it is
gratifying that relation (\ref{32}), which is a highly non-trivial
consequence of the dynamics of the system, is so well fulfilled.

The scale $\mu$ at which the MSV works was found to be the one where
$\al_s(\mu )=0.57$ in \ct{N3} by using the energy sum-rule. With the
help of the action sum-rule we now checked the consistency of the MSV
at this scale and obtained an effective $\tilde{\be}$-function.

As can be seen from \eq{11} and \eq{12} the squared field strengths
$\vec{B}^2$ and $\vec{E}^2$ depend on the renormalization scale $\mu$
and hence on $\al_s(\mu)$. We may use relations (\ref{12}) and
(\ref{31}) in order to predict the ratio of
\[Q\equiv \frac{\int {\rm d^3}x <\vec B(x)^2>_R}{\int {\rm d^3}x<\vec
E(x)^2>_R}= \frac{2+\tilde{\be}/\al_s}{2-\tilde{\be}/\al_s} \]
as a function of the strong coupling $\al_s$. The result is plotted in
\fig{BOVERE} (we have used there our numerically determined value of
$\be_2$). This ratio can be checked in lattice gauge calculations.

\epsfxsize7cm
\begin{figure}[ht]
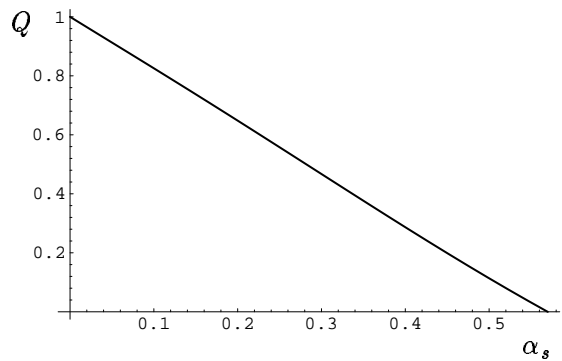
\caption{The ratio $Q$ as a function of $\al_s$}
\lbfi{BOVERE}

In this paper we have worked in pure gauge theory and in a theory with
gluons and one heavy quark field. We leave the investigation of
similar sum-rules in a theory including light quarks for future work.

{\Large \bf Acknowledgments} The authors thank E. Laermann, C.
Michael, H.J. Rothe and I. O. Stamatescu for discussions and
especially M. Jamin for valuable comments and suggestions.

\end{document}